\documentclass[12pt]{article}

\usepackage{newtxtext,newtxmath}
\usepackage{graphicx}

\usepackage[letterpaper,margin=1in]{geometry}

\renewenvironment{abstract}
	{\quotation}
	{\endquotation}

\date{}

\makeatletter
\renewcommand{\fnum@figure}{\textbf{Figure \thefigure}}
\renewcommand{\fnum@table}{\textbf{Table \thetable}}
\makeatother

\usepackage{scicite}

\usepackage{url}

\def\scititle{
	Loss-insensitive reduction of amplified quantum noise in a Raman amplifier with coherent feedback
}
\title{\bfseries \boldmath \scititle}

\author{
	Jianmin Wang$^{\ast}$,
	Rong Zhu$^{\ast}$,
	Z. Y. Ou$^{\dagger}$\and
	\small Department of Physics, City University of Hong Kong, 83 Tat Chee Avenue, Kowloon, Hong Kong.\and
	\small$^\dagger$Corresponding author. Email: jeffou@cityu.edu.hk\and
	\small$^\ast$These authors contributed equally to this work.
}

\begin{document} 

% Insert the title and author list
\maketitle

% Abstract, in bold
% There are strict length limits, and not all formats have abstracts.
% Consult the journal instructions to authors for details.
% Do not cite any references in the abstract.
\begin{abstract} \bfseries \boldmath
A quantum amplifier usually adds extra noise inevitably through coupling to internal degrees of freedom while amplifying the signal. The introduction of quantum correlations can effectively suppress this extra noise. In this work, we utilize the established quantum correlation between the Stokes field and atomic spin waves in the Raman amplification process to feed back a portion of the Stokes field into the amplifier. This leads to a reduction in amplified quantum noise that is independent of the feedback loss at high gain. A maximum of $6\,\mathrm{dB}$ noise reduction is observed. The single-path feedback amplifier is found to be sensitive to the feedback phase due to quantum interference, and we experimentally demonstrate an improvement in the signal-to-noise ratio of $4.4\pm 0.7\,\mathrm{dB}$ for a phase signal encoded in the feedback loop compared to the case without feedback. This general concept can be extended to integrated optics and fiber-optic systems with potential applications in quantum precision measurement.
\end{abstract}

% The first paragraph of any Science paper does NOT have a heading
% Nor is it indented

\noindent
Amplifiers are widely used instruments in modern science and technology to enhance small signals to a level that can be observed. In the process, noise is inevitably amplified as well. Down to the fundamental level, when all the classical noise is eliminated, quantum fluctuations play an important role in the noise performance of the amplifier\cite{hef62,hau62,hong85,gla86} and add excess noise in addition to the amplification of the input noise, leading to the infamous high-gain 3 dB degradation in the signal-to-noise ratio \cite{hong85}. 

The added extra noise is due to the coupling to the internal degrees of freedom of the amplifier to acquire gain for the input. These internal modes are usually unattended and are independent of the input. Their quantum fluctuations give rise to the extra noise. Parametric amplifiers are a special type of amplifier achieved through nonlinear parametric interaction processes in which an idler field is involved and serves as the internal mode of the amplifier. Due to its accessibility, the idler field can be manipulated to alter the noise performance of the amplifier. In particular, when the idler field is placed in a squeezed vacuum state, its quantum fluctuations can be suppressed, and the extra noise added to the amplifier can be reduced \cite{mil87,ou93}. The amplifier output noise can be further reduced when its internal modes are placed in a quantum-correlated state with the input field \cite{ou93b,kong13,wang25}. 

However, such approaches rely on the external preparation of squeezed or entangled states, which are vulnerable to optical loss during propagation and mode mismatch between the entanglement source and the amplifier. Even small losses can degrade quantum correlations exponentially, whereas mode mismatch introduces classical noise that is amplified alongside the signal, undermining the intended noise reduction. Given this, a coherent feedback architecture that combines correlation generation and amplification in one device provides a simpler and more robust alternative to the traditional two-component setup\cite{seth00,gou09}. Coherent feedback has emerged as a versatile tool in quantum optics over the past decade, enabling dynamic control of quantum systems without relying on external measurement or state preparation, and has been adopted so far to assist with a diverse range of tasks\cite{say11,zhong20,har20,ern23,vod24}. By recycling the amplifier’s own output to establish quantum correlations internally, this approach eliminates the need for external entanglement sources and the associated mode matching. 

So far, most theoretical and experimental studies on noise reduction via entanglement or input-state correlation have been carried out in optical parametric amplifiers\cite{ou93,kong13}, where the internal idler optical modes can be easily optically coupled from the outside. Even coherent feedback implementations for noise reduction are predominantly based on such optical amplifiers\cite{yok20,san12}. In contrast, for amplifiers using atoms as the gain medium, internal atomic degrees are generally not directly accessible. This inaccessibility has hindered the realization of noise reduction in such systems until recently, with demonstrations relying on dual Raman processes within the same atomic ensemble\cite{wang25}. Notably, the localization of internal atomic degrees in Raman amplifiers was once a challenge for noise reduction via entanglement or input-state correlation, as internal atomic degrees of freedom are difficult to utilize. This localization, however, turns out to be an advantage in coherent feedback schemes.

\begin{figure}
\centering
\includegraphics[width=0.8\columnwidth]{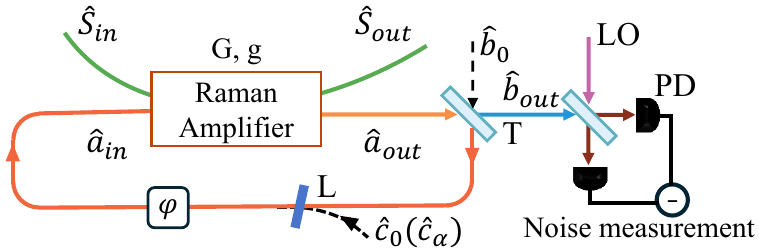}
	\caption{\textbf{Conceptual schematic for a Raman amplifier with single-path feedback.}
    The Raman amplifier, labeled with its gain coefficients $G$ and $g$ which satisfy the relation $G^2-g^2=1$, amplifies the input signal field and couples the amplified field $\hat{a}_{\mathrm{out}}$ into the feedback loop, which feeds back to the amplifier’s input as $\hat{a}_{\mathrm{in}}$; a beam splitter with transmittance $T$ then splits the output field, routing a portion $\hat{b}_{\mathrm{out}}$ to a photodetector (PD) that is combined with a local oscillator (LO) for noise measurement, while the remaining field continues in the feedback path. The orange closed loop forms the feedback channel, which introduces additional vacuum noise $\hat{c}_0$ via loss and adjusts the phase of the feedback loop using a piezoelectric transducer; for phase sensing, the vacuum input $\hat{c}_0$ is replaced by the coherent input field $\hat{c}_\alpha$ shown in parentheses. $\hat{b}_0$ represents the additional vacuum noise introduced by the feedback loop. The green lines denote the input $\hat{S}_{\mathrm{in}}$ and output $\hat{S}_{\mathrm{out}}$ at the idler port of the Raman amplifier, which corresponds to the non-propagating atomic state.}
	\label{Fig1}
\end{figure}

In this paper, we investigate a coherent feedback amplifier, which utilizes the same Raman amplifier to prepare atoms in states correlated with the Stokes field by tapping a small portion of the output Stokes field and feeding it back to the amplifier, as seen in Fig.\ref{Fig1}. This arrangement allows the internal states of the atoms, which do not travel, to be correlated with the input and leads to a reduction in the noise of the amplifier’s output. We investigate in detail the quantum noise performance of the Raman amplifier through this mechanism under various feedback conditions. In particular, when  transmission is maximized and the loss is near zero, we achieve a $6\,\mathrm{dB}$ reduction of amplified quantum noise that exceeds the noise suppression level attainable using separate amplifiers. This single-amplifier coherent feedback configuration reduces system complexity, eliminates noise caused by mode mismatch, and provides quantum noise suppression that is insensitive to feedback loss at high gain. We further find that the amplifier output is sensitive to the feedback phase due to interference. By encoding a phase signal in the feedback path using an electro-optic modulator (EOM), we observe a signal-to-noise ratio improvement of $4.4 \pm 0.7~\mathrm{dB}$ compared with the no-feedback case under the same operating conditions. Atomic degrees of freedom are involved in this scheme, which can be used as a quantum sensor to detect changes in the atomic state.

\subsection*{Theoretical principles}
% {\it Theory --} 
According to the general theory of quantum amplifiers \cite{cav82}, the output mode is amplified from the input by an amplitude gain of $G$, but is also coupled to its internal modes. For a Raman amplifier with Alkaline atoms as medium, the internal mode is the atomic spin wave characterized by a Bosonian operator $\hat S$ and the input-output relation is given by
\begin{equation}\label{a-out1}
\hat{a}_{\text {out }}=G \hat{a}_{\mathrm{in}}+g \hat{S}_{\mathrm{in}}^{\dagger},
\end{equation}
where $G$ is the amplitude gain with $g=\sqrt{G^2-1}$, and $\hat{S}_{\mathrm{in}}$ satisfies $[\hat{S}_{\mathrm{in}},\hat{S}_{\mathrm{in}}^{\dagger}]=1$, relating to the atomic coherence. To achieve an appreciable gain for the amplifier, the atomic medium is driven by a strong classical field and $G$ is exponentially related to the pump power. Eq.(\ref{a-out1}) resembles the input-output relationship of a parametric amplifier.

Note that Eq.(\ref{a-out1}) is a relation for field operators and we can rewrite it in the quadrature-phase amplitude form as
\begin{equation}\label{X-out1}
\hat{X}_{{\text{out} }}=G \hat{X}_{a_{\mathrm{in}}}+g \hat{X}_{S_{\mathrm{in} }},
\end{equation}
where $\hat X_O \equiv \hat O+\hat O^{\dagger}$ with $\hat O = \hat a_{\mathrm{in},\mathrm{out}}, \hat S_{\mathrm{in}}$.
When the atomic medium is initially prepared in ground states and the input field is in vacuum or coherent state, $\hat{X}_{a_{\mathrm{in} }}$ and $\hat{X}_{S_{\mathrm{in} }}$ are independent of each other and we obtain the output noise level as 
\begin{equation}\label{Xqn}
\left\langle\Delta^2 X_{\text{out}} \right\rangle = G^2 + g^2 = 2G^2-1 \equiv G_{\mathrm{qn}}.
\end{equation}
This is typical of amplifier noise consisting of amplified input vacuum noise ($G^2$) and extra noise ($g^2$) added from the internal atomic mode $\hat S_{\mathrm{in}}$. We define this as the quantum noise gain $G_{\mathrm{qn}}$ by the amplification of the vacuum noise for both the input and internal modes.

However, when atoms ($\hat S_{\mathrm{in}}$) are prepared in a correlated state with the input field ($\hat a_{\mathrm{in}}$), the fluctuation of $\hat{X}_{a_{\mathrm{in} }} + \hat{X}_{S_{\mathrm{in} }}$ can be smaller than their respective vacuum fluctuations \cite{rei89,ou92}, leading to noise reduction \cite{kong13}. 
 Our approach is to tap a small portion of the output Stokes and feed it back to the amplifier for noise cancellation by quantum correlation.

As shown in Fig.\ref{Fig1}, we use a beam splitter with transmissivity $T$ to split a part of the output Stokes field and feed it back to the amplifier. An extra phase shift $\varphi$ is added to control the feed-back phase. We also add loss $L$  using a beam splitter with transmissivity $1-L$ to couple in the vacuum mode $\hat{c}_{0}$: 
\begin{equation}\label{a-in}
\hat{a}_{\text{in}}=e^{i \varphi}\sqrt{1-L}\left( \sqrt{T}\hat{b}_0-\sqrt{1-T}\hat{a}_{\text{out}} \right)+e^{i \varphi}\sqrt{L} \hat{c}_0,
\end{equation}
where $\hat b_0$ is the input mode. Substituting Eq.~(\ref{a-in}) into Eq.~(\ref{a-out1}), we obtain
\begin{equation}\label{a-out2}
\begin{split}
\hat{a}_{\text{out}} &= -G e^{i \varphi} \sqrt{(1-T)(1-L)} \hat{a}_{\text{out}} \\
&\quad + G e^{i \varphi} \sqrt{(1-L) T} \hat{b}_0 + e^{i \varphi}G \sqrt{L} \hat{c}_0 + g \hat{S}_{\text{in}}^{\dagger}
\end{split}
\end{equation}
or
\begin{equation}\label{a-out3}
\hat{a}_{\text{out}} = \frac{G e^{i \varphi} \sqrt{(1-L) T} \hat{b}_0  + e^{i \varphi}G \sqrt{L} \hat{c}_0 + g \hat{S}_{\mathrm{in}}^{\dagger}}{1 + G e^{i \varphi} \sqrt{(1-T)(1-L)}}.
\end{equation}
For the output port, we have $\hat{b}_{\text{out}}=\sqrt{T}\hat{a}_{\text{out}}+\sqrt{1-T}\hat{b}_{\text{0}}$, and according to Eq.(\ref{a-out3}), we obtain
\begin{equation}\label{b-out}
\begin{split}
\hat{b}_{\text{out}} &= \frac{ \left( \sqrt{1-T} + G e^{i \varphi} \sqrt{1-L} \right)}{1 + G e^{i \varphi} \sqrt{(1-T)(1-L)}}\hat{b}_0 \\
&\quad + \frac{e^{i \varphi}G \sqrt{T L} \hat{c}_0 + g\sqrt{T} \hat{S}_{\mathrm{in}}^{\dagger}}{1 + G e^{i \varphi} \sqrt{(1-T)(1-L)}}.
\end{split}
\end{equation}

We measure the noise variance of the fields by homodyne detection of the quadrature-phase amplitude $\hat X = \hat a+\hat a^{\dagger}$. To see the effect of noise performance with feedback, we compare it to the case when there is no coherent feedback, that is, $L=1$ or the coherent feedback is completely lost. The noise variance of the quadrature amplitude at the output port can be calculated from Eq.(\ref{b-out}) by setting $L=1$ and is
\begin{equation}\label{Xa}
\left\langle\Delta^2 X_{b}^T \right\rangle = T(2G^2-1) +1-T,
\end{equation}
with $b\equiv b_{\mathrm{out}}$. This is just the noise level of the output of a regular amplifier after passing through a beam splitter with transmissivity $T$. This sets the reference noise level that we will compare to for noise reduction in the feedback case.

When $T \neq 1$, i.e., a portion of the Stokes field is fed back into the Raman amplifier with correlated atomic spin waves, the output is phase-sensitive, especially when $e^{i \varphi} =-1$ and the denominator becomes zero, or a threshold of oscillation is achieved when $G_{\mathrm{th}} = 1/\sqrt{(1-T)(1-L)}$. On the other hand, the absolute value of the denominator becomes maximum when $e^{i \varphi} =1$, which minimizes the contribution of the last two terms in Eq.(\ref{b-out}) that are the sources of the extra noise. In this case, the noise variance of output port $\hat b_{\mathrm{out}}$ is
\begin{equation}\label{Xb}
\begin{aligned}
\left\langle\Delta^2 X_b\right\rangle & =\frac{\left(\sqrt{1-T}+G \sqrt{1-L} \right)^2+G^2 T L+g^2 T}{\left(1+G  \sqrt{(1-T)(1-L)}\right)^2} \\
& =\frac{T\left(G^2+g^2-1\right)}{\left(1+G /G_{\mathrm{th}}\right)^2}+1,
\end{aligned}
\end{equation}
with $b\equiv b_{\mathrm{out}}$. By comparing Eq.(\ref{Xa}) and Eq.(\ref{Xb}), we obtain the noise reduction factor
\begin{equation}\label{R}
\begin{aligned}
R & \equiv \frac{\left\langle\Delta^2 X_b\right\rangle}{\left\langle\Delta^2 X_b^T\right\rangle} \\
&=\frac{2T\left(G^2-1\right)+\left(1+G /G_{\mathrm{th}}\right)^2}{\left(1+G  /G_{\mathrm{th}}\right)^2(2TG^2+1-2T)}.
\end{aligned}
\end{equation}
The noise reduction is characterized by $R < 1$, which depends on the parameters of $T, L$, and $G$. 

To see this, let us first look at the special case of $T\approx 1$. In this case, Eq.(\ref{R}) becomes
\begin{equation}\label{R2}
R \approx \frac{2G^2-2+\left(1+G/G_{\mathrm{th}}\right)^2}{\left(1+G/G_{\mathrm{th}}\right)^2(2G^2-1)},
\end{equation}
which approaches 1/4 or $-6\,\mathrm{dB}$ noise reduction when $G\rightarrow G_{\mathrm{th}}\approx 1/\sqrt{(1-T)(1-L)} \gg 1$. Here, the limit is taken with $G/G_{\mathrm{th}}\rightarrow1$ as $T\rightarrow1$, so that the increasing amplifier gain compensates for the decreasing feedback fraction. This is amazing considering that a relatively large amount of noise reduction $-6\,\mathrm{dB}$ is achieved with feedback of only a small fraction ($1-T\ll1$) of the output.

For other values of $T$, we find from Eq.(\ref{R}) that $R\approx 1/(1+G/G_{\mathrm{th}})^2$ when $TG^2\gg 1$ and also approaches $1/4$ for $G\rightarrow G_{\mathrm{th}}$, and this is so regardless of the magnitude of the feedback loss $L$. So, the noise reduction factor is loss-insensitive, in contrast to the general belief that losses always degrade the reduction of amplified vacuum noise. The reason for this is that the feedback part does not create the quantum correlation but is simply an injection into the amplifier, which is where the quantum correlation is generated for noise reduction. Although the above analysis focuses on the $\hat{X}$ quadrature, the orthogonal phase quadrature $\hat{Y}=i(\hat{a}^{\dagger}-\hat{a})$ has the same noise variance, $\langle\Delta^{2}\hat{Y}\rangle=\langle\Delta^{2}\hat{X}\rangle=\langle\Delta^{2}\hat{X}(\theta)\rangle$, which means that the noise is minimum for all quadrature-phase amplitudes $\hat{X}(\theta)$. This indicates that the underlying physics for noise reduction here is quantum destructive interference, which suppresses fluctuations of the entire field and hence reduces noise in all quadrature-phase amplitudes, similar to SU(1,1) interferometers \cite{yur86,kong13}.

The noise reduction property described above in the phase-sensitive amplification with feedback can find practical applications in phase sensing. To quantify its effect on the signal-to-noise ratio (SNR), we first consider a single Raman amplifier without feedback as a reference. A coherent field carrying a small phase signal $\delta$ is injected into the amplifier, with $\langle\hat{a}_{\mathrm{in}}\rangle=\alpha_\mathrm{NF} e^{i\delta}$, where $\alpha_\mathrm{NF}$ is real and $|\delta|\ll1$. According to Eq.~(\ref{a-out1}), the mean output field is $\langle\hat{a}_{\mathrm{out}}\rangle=G\alpha_\mathrm{NF} e^{i\delta}$. The signal measured in the phase quadrature $\hat{Y}_{\mathrm{NF}}=i(\hat{a}_{\mathrm{out}}^\dagger-\hat{a}_{\mathrm{out}})$ is therefore
\begin{equation}\label{eq:signal_noFB}
\langle\hat{Y}_{\mathrm{NF}}\rangle^2
\simeq4G^2\alpha_\mathrm{NF}^2\delta^2.
\end{equation}
The output phase-quadrature noise is given by Eq.~(\ref{Xqn}), namely $\langle\Delta^2\hat{Y}_{\mathrm{NF}}\rangle=2G^2-1$. The resulting $\mathrm{SNR}_{\mathrm{NF}}$ is
\begin{equation}\label{eq:SNR_noFB}
\begin{aligned}
\mathrm{SNR}_{\mathrm{NF}}
&\equiv
\frac{\langle\hat{Y}_{\mathrm{NF}}\rangle^2}
{\langle\Delta^2\hat{Y}_{\mathrm{NF}}\rangle} \\
&\simeq
\frac{4G^2 I^{\mathrm{NF}}_{\mathrm{ps}}\delta^2}{2G^2-1}
\simeq 2I^{\mathrm{NF}}_{\mathrm{ps}}\delta^2,
\end{aligned}
\end{equation}
where $I^{\mathrm{NF}}_{\mathrm{ps}}=\alpha_\mathrm{NF}^2$ is the phase-sensing intensity, and the last approximation holds in the high-gain regime $G^2\gg1$. We next analyze how the noise reduction translates into an improved SNR. With a coherent field injected through the $\hat{c}_0$ port, we rewrite the output field $\hat{b}_{\mathrm{out}}$ as
\begin{equation}\label{b-out_coh}
\begin{split}
\hat{b}_{\text{out}} &= \frac{ \left( \sqrt{1-T} + G e^{i \varphi} \sqrt{1-L} \right)}{1 + G e^{i \varphi} \sqrt{(1-T)(1-L)}}\hat{b}_0 \\
&\quad + \frac{e^{i \varphi}G \sqrt{T L} \hat{c}_{\alpha} + g\sqrt{T} \hat{S}_{\mathrm{in}}^{\dagger}}{1 + G e^{i \varphi} \sqrt{(1-T)(1-L)}},
\end{split}
\end{equation}
where $\hat{c}_{\alpha}$ describes the injected coherent field with coherent amplitude $\langle\hat{c}_{\alpha}\rangle=|\alpha|$. We consider a small phase excursion $\varphi=\delta$, with
$|\delta|\ll1$, about the noise-suppression operating point
$\varphi=0$. The signal, defined as the squared mean of the phase quadrature $\hat{Y}_{\mathrm{FB}}=i(\hat{b}_{\mathrm{out}}^\dagger-\hat{b}_{\mathrm{out}})$, is given by:
\begin{equation}\label{eq:Yb_expect}
\begin{split}
\langle\hat{Y}_{\mathrm{FB}}\rangle^{2} &= \langle i(\hat{b}^\dagger_{\rm out} - \hat{b}_{\rm out}) \rangle^2  \\
&= \frac{4G^2TL\alpha^2\delta^2}{\left(1 + G\sqrt{(1-T)(1-L)} \right)^2}. 
\end{split}
\end{equation}
The phase shift $\delta$ is mapped onto an amplified displacement of the output phase quadrature, with the response rescaled by coherent feedback. It is known that phase sensing signal highly depends on the intensity of the field that probes the phase change. To have a fair comparison with the scheme without feedback, we next calculate the intensity $I^{\mathrm{FB}}_{\mathrm{ps}}=\langle\hat{a}_{\mathrm{in}}^{\dagger}\hat{a}_{\mathrm{in}}\rangle$ used for phase sensing here. Substituting Eq.(\ref{a-out3}) into Eq.(\ref{a-in}) and simplifying, we obtain:
\begin{align}
\hat{a}_{\mathrm{in}} &=
\frac{e^{i\varphi}\sqrt{(1-L)T}\,\hat{b}_0
+e^{i\varphi}\sqrt{L}\hat{c}_\alpha
-g e^{i\varphi}\sqrt{(1-L)(1-T)}\,\hat{S}_{\mathrm{in}}^\dagger}
{1+G e^{i\varphi}\sqrt{(1-T)(1-L)}}. 
\end{align}
At the noise-suppression operating point $\varphi=0$, the photon number of the phase sensing field is given by 
\begin{align}
I^{\mathrm{FB}}_{\mathrm{ps}}&= \frac{L\alpha^2}{\left(1 + G\sqrt{(1-T)(1-L)} \right)^2}. 
\label{Ips}
\end{align}
With the output noise given in Eq.(\ref{Xb}), we obtain the SNR at the output as
\begin{equation}\label{eq:SNR_feedback}
\begin{split}
\mathrm{SNR}_{\mathrm{FB}} &\equiv
\frac{\langle\hat{Y}_{\mathrm{FB}}\rangle^2}
{\langle\Delta^2\hat{Y}_{\mathrm{FB}}\rangle} \\
&\simeq 2{\left(1 + G\sqrt{(1-T)(1-L)} \right)^2}I^{\mathrm{FB}}_{\mathrm{ps}}\delta^2. 
\end{split}
\end{equation}
By comparing $\mathrm{SNR}_{\mathrm{FB}}$ and $\mathrm{SNR}_{\mathrm{NF}}$ at matched phase-sensing intensities, $I_{\mathrm{ps}}^{\mathrm{NF}}=I_{\mathrm{ps}}^{\mathrm{FB}}$, we obtain a feedback-induced SNR enhancement factor of $\left(1+G\sqrt{(1-T)(1-L)}\right)^2$, which depends on the amplifier gain $G$ and the feedback parameters $T$ and $L$. The underlying physical picture can be understood from Eqs.~(\ref{Ips}) and (\ref{eq:SNR_feedback}), which show that, through destructive interference, coherent feedback reduces the circulating phase-sensing field while also suppressing the noise accompanying amplification, resulting in a lower noise background at the same $I_{\mathrm{ps}}$. This scheme is useful when a component in the feedback loop imposes a limit on the optical intensity, allowing a higher SNR to be achieved under the same constraint.

\subsection*{Experimental results and analysis}
\begin{figure}
\centering
\includegraphics[width=0.9\columnwidth]{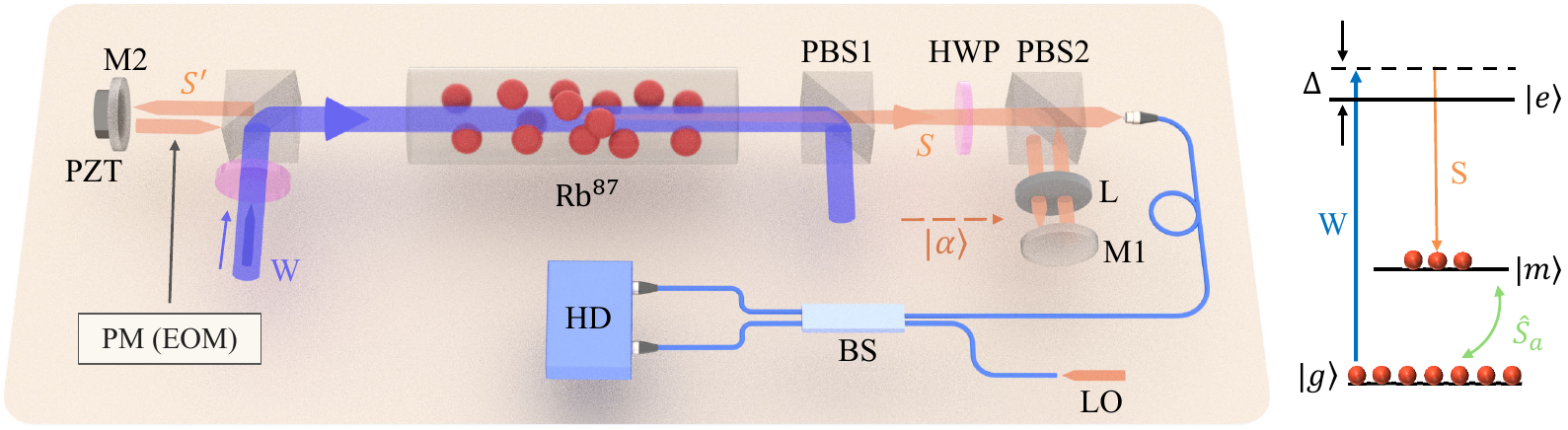}
	\caption{\textbf{Schematic of the coherent feedback Raman amplifier setup based on a ${}^{87}\mathrm{Rb}$ atomic ensemble.} A pump field (W) propagates through the atomic ensemble, driving the Raman interaction by optically pumping atoms from the ground state $|g\rangle$ to the excited state $|e\rangle$. This process generates a Stokes field ($S$) via stimulated Raman scattering, as atoms relax from $|e\rangle$ to a metastable state $|m\rangle$ and coherently excite a spin wave ($\hat S_{a}$) within the ensemble. A portion of the Stokes output field is directed into a coherent feedback loop via a polarization beam splitter (PBS) and mirror (M1). The feedback loop includes an attenuator (L) to control feedback loss, and a combination of a half-wave plate (HWP) and PBS that sets the signal transmissivity (T). The field is then retroreflected as the feedback Stokes field ($S^\prime$) back into the atomic ensemble via a piezo-electric transducer (PZT) mounted on mirror (M2) for phase scanning. For phase-sensing measurements, a coherent Stokes field (dashed line) is injected into the feedback loop through an additional beam splitter (not shown) placed before M1. An electro-optic modulator (EOM) is inserted before M2 to apply phase modulation (PM), with its insertion point indicated by the arrow. A local oscillator (LO) is combined with the output field at a beam splitter (BS) for homodyne detection (HD). Unwanted light is routed to a beam dump (BD).}

    %M: mirror; PBS: polarization beam splitter; PZT: piezo-electric transducer; HWP: half wave plate; $S$: Stokes field; $S^\prime$: Stokes field entering the amplifier via the feedback loop; L: Feedback Loss, controlled by attenuator; T: transmissivity of signal, controlled by HWP and PBS; LO: local oscillator; PD: photo detector; ESA: Electrical Spectrum Analyzer; BD: Beam dump.} 
	\label{setup}
\end{figure}

We next present experimental verification of the theory. The experimental schematic is shown in Fig.\ref{setup}. A cylindrical Pyrex cell with a length of 75~mm contains pure Rb-87 atomic vapor serving as the Raman amplifier (RA). The cell is placed inside a four-layer magnetic shield for isolating the magnetic field of the Earth and is heated up to $70~^\circ \mathrm{C}$ by a heating belt. The relevant atomic levels are shown in the inset of Fig.~\ref{setup}. Optical pumping prepares the atomic ensemble in the ground state $|g\rangle=|5^2S_{1/2},F=1\rangle$. A continuous-wave pump field $W$, derived from a single-frequency laser and detuned by $\Delta=800~\mathrm{MHz}$ from the $|g\rangle\rightarrow|e\rangle$ transition, drives the Raman interaction, where $|e\rangle=|5^2P_{1/2},F=2\rangle$. This process generates the Stokes field $S$ together with a collective atomic spin excitation between the two hyperfine ground states $|g\rangle$ and $|m\rangle=|5^2S_{1/2},F=2\rangle$.

The Stokes and pump fields have orthogonal polarizations and are separated and recombined using polarizing beam splitters. A portion of the Stokes output is extracted for detection, while the returned field $S^\prime$ is redirected to the same RA through mirrors M1 and M2, forming the coherent-feedback loop. Blocking M1 corresponds to no-feedback case. The output Stokes field is coupled into an optical fiber with a power coupling efficiency of approximately $70\%$ and measured using balanced homodyne detection with a balanced photodetector (Thorlabs, PDB435A). For phase sensing, a coherent Stokes field is injected into the feedback loop and combined with the returned field $S^\prime$ at a beam splitter (not shown in the figure) placed before mirror M1. The homodyne signal is analyzed using an electronic spectrum analyzer (ESA; Keysight, N9010B), with the resolution bandwidth (RBW) and video bandwidth (VBW) set to 1 MHz and 1 kHz, respectively.

The Raman amplifier whose noise behavior we aim to characterize is pumped by the continuous light field W, and we measure its output quantum noise level by homodyne detection with its local oscillator (LO) derived from another laser that is frequency tuned close to the Stokes field (about 6.8~GHz below the pump). Normally, the input field of the Raman amplifier is independent of the atomic states. To generate correlations between the atomic states and the input field and avoid mode mismatch, we use the output field $S$ of the RA, which is now correlated with the atomic states, as the input signal to the RA by reflecting it back via mirrors and reinjecting it to interact with the same ensemble of atoms. We investigate the effect of $S^\prime$ on the quantum noise level of the RA.

\begin{figure}
\centering
\includegraphics[width=0.8\columnwidth]{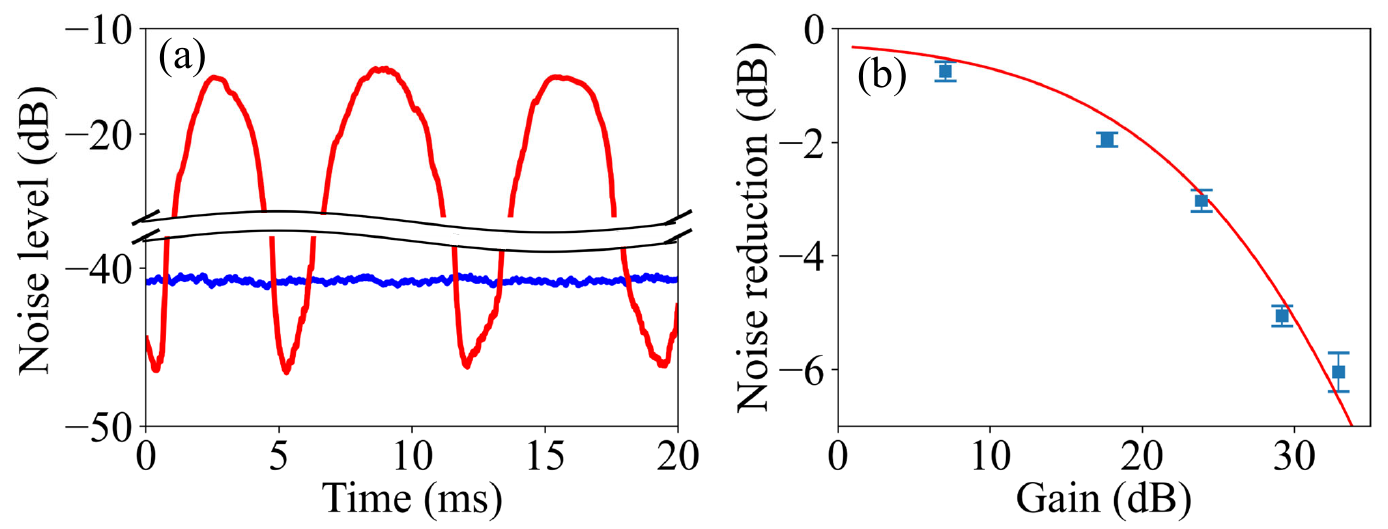}
	\caption{\textbf{Noise reduction of a Raman amplifier.} (a) Quantum noise level of the output as the feedback phase is scanned. The vacuum or shot noise level is $-73.6\,\mathrm{dB}$. (b) Noise reduction ratio as a function of the quantum gain $G_{\mathrm{qn}}$ of Raman amplifier without feedback. The solid curve is a fit to Eq.(\ref{R2}).} 
	\label{Fig-scan}
\end{figure}

As a reference for comparison, we record the noise level of the output Stokes field $S$ of the RA without the injection of $S^\prime$. This corresponds to a case where the input field is independent of the atomic medium, and the measured noise level is simply the amplified vacuum noise level. Next, we consider the case of $T\approx 1$ and the feedback is achieved with a mirror $\mathrm{M1}$ and a nearly transparent reflective surface (in the experiment, it is merely the surface reflection of the detector). Subsequently, we quantified the output noise level without feedback by blocking $\mathrm{M1}$, using this as the noise reference level, and Fig.~\ref{Fig-scan}(a) shows the result for $G_{\mathrm{qn}} = 33\,\mathrm{dB}$, which is close to the threshold of oscillation. The blue trace is the quantum noise level of the amplifier without feedback. Then we unblocked the feedback loop to generate a quantum correlation between the atomic medium and the input field to the RA. We scan the phase of the reinjected Stokes field $S^\prime$ with a piezoelectric transducer. The measured output Stokes noise level is shown in Fig.~\ref{Fig-scan}(a) as the red trace when the feedback is on and the feedback phase $\varphi$ is scanned in time. $6\,\mathrm{dB}$ of noise reduction is observed. We measure the amount of noise reduction as a function of the quantum gain of the bare amplifier, which increases with the pump power, and plot the results in Fig.\ref{Fig-scan}(b). As can be seen, the noise reduction increases with the increase of quantum gain of the bare amplifier, which is consistent with the theoretical predictions. The solid red curve is a fit from Eq.~(\ref{R2}).

\begin{figure}
\centering
\includegraphics[width=0.8\columnwidth]{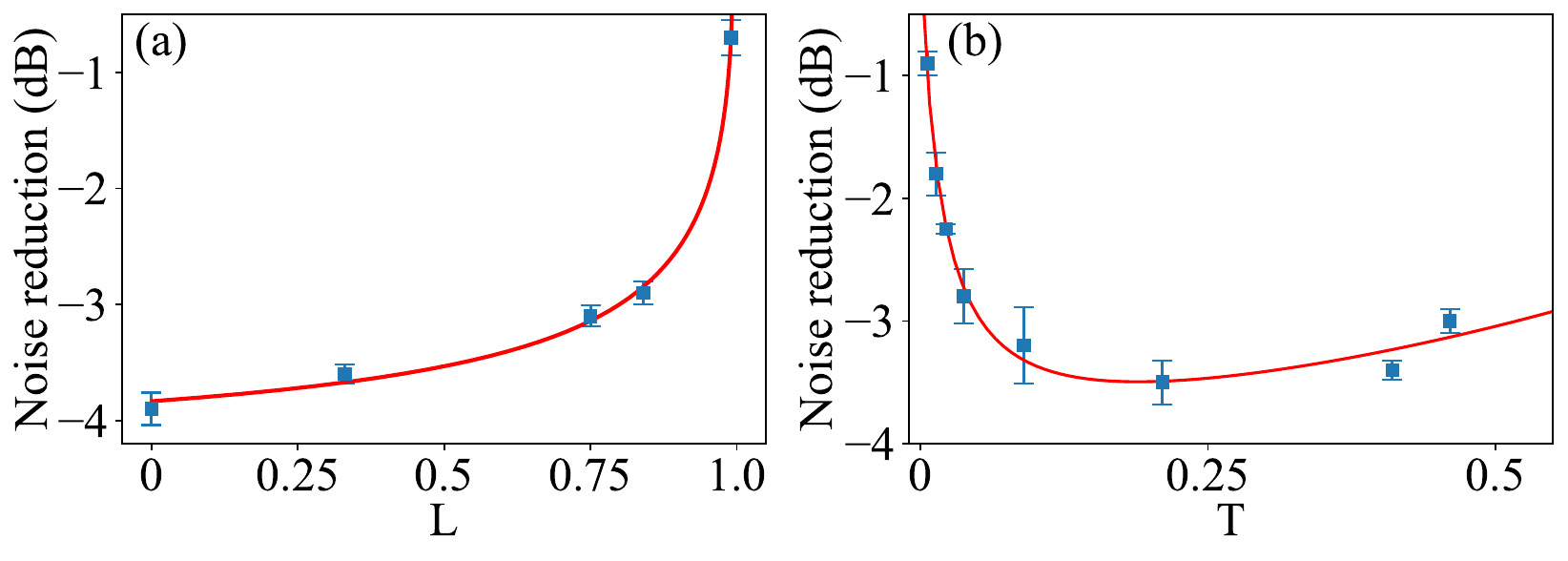}
	\caption{\textbf{Noise reduction as a function of feedback parameters with theoretical fitting.} (a) Measured noise reduction as a function of feedback loss $L$, with the solid red curve representing a theoretical fit from Eq.~(\ref{R}). (b) Measured noise reduction as a function of beam splitter transmittance $T$, with the solid red curve representing a theoretical fit from Eq.~(\ref{R}).} 
	\label{Fig3}
\end{figure}

Next, to investigate the impact of feedback parameters $T$ and $L$ on noise reduction, we replace the nearly transparent surface with a combination of polarization beam splitter, a half-wave plate and mirror $\mathrm{M1}$ for variable feedback $T$ and loss $L$, as shown in Fig.\ref{setup}. The parameters $T$ and $L$ affect $G_{\mathrm{th}}$ since $G_{\mathrm{th}}\approx 1/\sqrt{(1-T)(1-L)}$. We first fix $T$ at 0.25, $G_{\mathrm{qn}} = 18$ dB and vary the feedback loss $L$ by attenuator, with the results shown in Fig.\ref{Fig3}(a). Then we fix $L$ at 0.01, $G_{\mathrm{qn}}= 23$ dB, and change $T$ by rotating the half-wave plate (HWP), with the results shown in Fig.~\ref{Fig3}(b). The solid red curves in Fig.~\ref{Fig3} are a fit to Eq.(\ref{R}). 

Fig.~\ref{Fig3}(a) shows that the single-path feedback is not strongly sensitive to feedback loss; even when the loss reaches $6\,\mathrm{dB}$ ($L=0.75$), the noise reduction can still reach $3\,\mathrm{dB}$. Fig.\ref{Fig3}(b) shows that there exists an optimal $T$ that allows a balance of the newly generated and feedback Stokes fields to reach an optimal noise reduction at finite gain. 

\begin{figure}
\centering
\includegraphics[width=0.5\columnwidth]{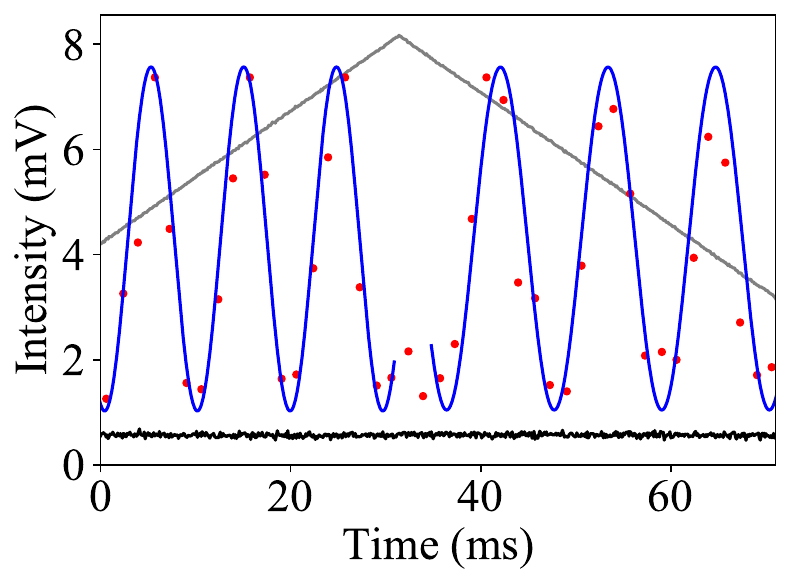}
	\caption{\textbf{Observed single-path feedback amplifier sensitive to feedback phase.} Interference fringes, represented by red data points with a blue sinusoidal fitting curve, are measured at the output of RA as a function of phase scan (gray). The black curve corresponds to the background intensity level.} 
	\label{Fig5}
\end{figure}

When a seed Stokes field is injected into the system, this feedback-based setup exhibits a phase-dependent interference response. The seed enters through a beam splitter (not shown in the figure) placed before mirror M1 in Fig.~\ref{setup}. Figure \ref{Fig5} shows the interference fringes as the phase of the back-injected Stokes field $S^\prime$ is scanned.

\begin{figure}
\centering
\includegraphics[width=1.0\columnwidth]{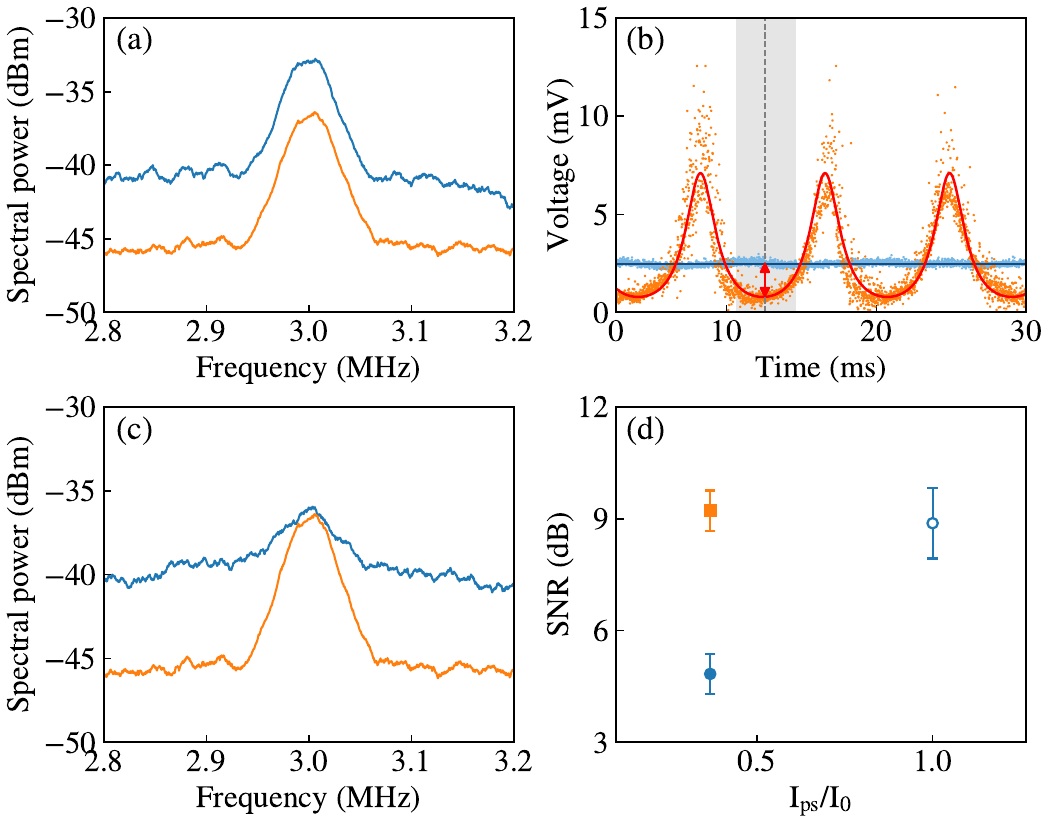}
	\caption{\textbf{Feedback-enhanced signal detection.} Orange and blue denote measurements with and without coherent feedback, respectively. (a) Measured output spectra for a phase signal at $3~\mathrm{MHz}$ encoded in the feedback loop. (b) Sensing intensity $I_{\mathrm{ps}}$, monitored through the photodetector voltage, during a scan of the feedback phase. The red solid curve is a fit to the feedback data, and the vertical dashed line marks the dark-fringe operating point. (c) Output spectra after reducing the sensing intensity in the no-feedback configuration to match that of the feedback configuration at the dark fringe. (d) SNR versus normalized sensing intensity $I_{\mathrm{ps}}/I_0$ for three configurations: with feedback at the dark fringe (orange square), without feedback at the original intensity (open blue circle), and without feedback at the reduced intensity (filled blue circle). Here, $I_0$ denotes the initial intensity of the coherent input field. At the same sensing intensity, coherent feedback yields a higher SNR.}
	\label{Fig6}
\end{figure}

For practical sensing applications, a small phase signal at $3~\mathrm{MHz}$ is encoded in the feedback loop using an EOM and measured by homodyne detection. As a reference, we first record the output spectrum without feedback and then repeat the measurement with feedback, as shown by the blue and orange traces, respectively, in Fig.~\ref{Fig6}(a). With feedback, the phase is locked at the dark fringe, yielding a noise reduction of $4.7\pm0.7~\mathrm{dB}$ relative to the no-feedback case. The corresponding SNRs are $8.9\pm0.9~\mathrm{dB}$ without feedback and $9.2\pm0.5~\mathrm{dB}$ with feedback, as indicated by the open blue circle and orange square, respectively, in Fig.~\ref{Fig6}(d). The two cases show similar SNRs, but this does not rule out an SNR enhancement, since coherent feedback changes the phase-sensing field. We will compare the SNRs at the same $I_{\mathrm{ps}}$.

To characterize this effect, we monitor the sensing intensity $I_{\mathrm{ps}}$ through the photodetector while scanning the feedback phase, as shown in Fig.~\ref{Fig6}(b). The relative values of $I_{\mathrm{ps}}$ are inferred from the output intensity. Without feedback, $I_{\mathrm{ps}}$ remains independent of the feedback phase, whereas with feedback it varies and reaches a minimum at the dark fringe. The red arrow in Fig.~\ref{Fig6}(b) indicates a reduction of $4.66\pm0.82~\mathrm{dB}$ in $I_{\mathrm{ps}}$ compared to the case without feedback, and the orange data are fitted using the general phase-dependent expression that reduces to Eq.~(\ref{Ips}) at $\varphi=0$, as shown by the solid red curve. Thus, the same coherent injection gives different sensing intensities in the two cases.

Based on this, we reduce the coherent input amplitude in the no-feedback case to match the phase-sensing intensity $I_{\mathrm{ps}}$ in the feedback case. The resulting spectra are shown in Fig.~\ref{Fig6}(c), where the blue and orange curves correspond to the cases without and with feedback, respectively. For a coherent input, lowering its amplitude reduces the phase signal but leaves its vacuum fluctuations unchanged. Consequently, the noise of the bare Raman amplifier remains governed by the amplified input vacuum noise, as described by Eq.~(\ref{Xa}). In the feedback configuration, however, the correlations between the returned Stokes field and the atomic spin wave suppress their combined contribution to the output noise. At the same $I_{\mathrm{ps}}$, the feedback amplifier therefore provides a signal readout with a lower noise background and a higher SNR.

Figure~\ref{Fig6}(d) plots the measured SNRs with and without feedback against the normalized sensing intensity $I_{\mathrm{ps}}/I_0$, where $I_0$ denotes the initial intensity of the coherent input field. In matched $I_{\mathrm{ps}}$, coherent feedback provides an SNR improvement of $4.4\pm0.7~\mathrm{dB}$ by reusing atom–light correlations within a single Raman amplifier, with readout through the Stokes field alone. This scheme is useful when a component in the feedback loop imposes an optical-intensity limit, allowing a higher SNR to be achieved under the same constraint.

In conclusion, we have experimentally shown a $6\,\mathrm{dB}$ reduction of the amplifier’s output noise in an atomic Raman amplifier. The experimental results were predicted by a detailed theoretical analysis and are in good agreement with them. Returning a fraction of the output Stokes field to the same atomic ensemble correlates the optical input with the internal atomic spin wave, enabling quantum destructive interference between their fluctuations during amplification. The resulting noise reduction is insensitive to feedback loss at high gain. For a phase signal encoded in the feedback loop, we further demonstrate an SNR improvement of $4.4\pm0.7~\mathrm{dB}$ at the same phase-sensing intensity. Together, these results provide theoretical and experimental insight into how coherent feedback suppresses amplification noise and improves signal readout in a single Raman amplifier. The required correlations are generated within the amplifier, and the signal is read out through the Stokes field alone, avoiding an external entanglement source and the associated mode matching. This concept is useful when a component in the feedback loop imposes an optical intensity limit. The coherent-feedback principle may also be extended to integrated photonic platforms, while the coupling between the optical field and the atomic spin wave suggests future applications in sensing perturbations to the atomic medium.

% Research Articles and Reviews split the text into sections using headings
% Use a short (up 6 words) descriptive phrase, not generic 'Results' or 'Conclusions'
% Most other formats do not have headings, see the journal instructions to authors for details

\clearpage % Clear all remaining figures and tables then start a new page


% If your text is very short you might need to uncomment the following line to avoid
% layout problems with the figures and tables.
%\newpage

%%%%%%%%%%%%%%%% MAIN TEXT FIGURES %%%%%%%%%%%%%%%

%%%%%%%%%%%%%%%% REFERENCES %%%%%%%%%%%%%%%

\clearpage % Clear all remaining figures and tables then start a new page

% The list of references goes after the main text and before the acknowledgements
% When preparing an initial submission, we recommend you use BibTeX, like this:
%
\bibliography{bibliography} % for a file named science_template.bib
\bibliographystyle{sciencemag}

% After the paper has completed peer review and been revised ready for acceptance,
% you should comment out the lines above and copy-paste the contents of your .bbl
% file here instead. This will help ensure that our conversion software works correctly.
% Remember to re-run BibTeX first - check the timestamp!
%
% Example of the first three entries copy-pasted from science_template.bbl:
%
%\begin{thebibliography}{1}
%
%\bibitem{example}
%A.~N. {Author}, An example reference. \emph{Journal of Improbable Research}
%  \textbf{1}, 67 (2020).
%
%\bibitem{example2}
%F.~M. {Surname}, S.~{Author}, A second example. \emph{Interesting Research
%  Letters} \textbf{32}, 897 (2019).
%
%\bibitem{example_preprint}
%P.~{One}, P.~{Two}, P.~{Three}, {An unpublished preprint}. \emph{preprint}
%  (2021), arXiv:2101.12345.
%
%\end{thebibliography}

%%%%%%%%%%%%%%%% ACKNOWLEDGEMENTS %%%%%%%%%%%%%%%

\section*{Acknowledgments}
\paragraph*{Funding:}
The work is supported by City University of Hong Kong (Project No.9610522) and the General Research Fund from Hong Kong Research Grants Council (Nos.11315822 and 11307625).
\paragraph*{Author contributions:}
Z. Y. Ou supervised the whole project. J. Wang, R. Zhu, and Z. Y. Ou conceived the research. J. Wang, R. Zhu, and Z. Y. Ou designed the experiments. J. Wang and R. Zhu performed the experiment. J. Wang and Z. Y. Ou contributed to the theoretical study. J. Wang, R. Zhu, and Z. Y. Ou analyzed the data. J. Wang drew the diagrams. J. Wang and Z. Y. Ou wrote the paper. All authors contributed to the discussion and review of the manuscript.
\paragraph*{Competing interests:}
There are no competing interests to declare.
\paragraph*{Data and materials availability:}
All data needed to evaluate the conclusions in the paper are present in the paper and/or the Supplementary Materials.

%%%%%%%%%%%%%%%% SUPPLEMENT LIST %%%%%%%%%%%%%%%

% List the contents of your Supplementary Materials, including the numbers of any
% supplementary figures, tables, external data files etc. and any references that are
% cited only in the supplement. In this example, refs. 7-8 are cited only in the supplement.
% Fill out your numbers accordingly and delete any lines that aren't applicable.

%%%%%%%%%%%%%%%% SUPPLEMENTARY REFERENCES %%%%%%%%%%%%%%%

% Do NOT include a reference list in the supplement.
% All references must be in a single list at the end of the main text.
% The copyeditors will ensure that the correct reference list appears with each version of the paper
% (print, HTML, PDF, mobile app, metadata for bibliographic databases etc.)

\end{document}